\documentclass[aps,prmaterials,reprint,superscriptaddress,amsmath,amssymb,floatfix,longbibliography,raggedbottom]{revtex4-2}

\usepackage{graphicx}
\usepackage{booktabs}
\usepackage{dcolumn}
\usepackage{bm}
\usepackage{placeins}
\usepackage[hidelinks]{hyperref}

\begin{document}
\graphicspath{{figures/}}

\makeatletter
\let\ps@manuscriptstyle\ps@plain
\makeatother
\setlength{\textheight}{\dimexpr\textheight-15pt\relax}
\setlength{\footskip}{45pt}
\pagestyle{manuscriptstyle}

\title{Electrostatic screening and nitrogen-vacancy charge states in boron-doped/nitrogen-doped diamond junctions}

\author{Masato Koga}
\affiliation{School of Engineering, Institute of Science Tokyo, 4259 Nagatsuta-cho, Midori-ku, Yokohama,
Kanagawa 226-8501, Japan}
\author{Haizen Cho}
\affiliation{School of Engineering, Institute of Science Tokyo, 4259 Nagatsuta-cho, Midori-ku, Yokohama,
Kanagawa 226-8501, Japan}
\author{El Mustapha Mansouri}
\affiliation{School of Engineering, Institute of Science Tokyo, 4259 Nagatsuta-cho, Midori-ku, Yokohama,
Kanagawa 226-8501, Japan}
\author{Eikichi Kimura}
\affiliation{School of Engineering, Institute of Science Tokyo, 4259 Nagatsuta-cho, Midori-ku, Yokohama,
Kanagawa 226-8501, Japan}
\author{Ryo Matsumoto}
\affiliation{Research Center for Materials Nanoarchitectonics (MANA),
National Institute for Materials Science,
1-2-1 Sengen, Tsukuba, Ibaraki 305-0047, Japan}
\author{Yoshihiko Takano}
\affiliation{Research Center for Materials Nanoarchitectonics (MANA),
National Institute for Materials Science,
1-2-1 Sengen, Tsukuba, Ibaraki 305-0047, Japan}
\author{Keigo Arai}
\email{arai.k.835f@m.isct.ac.jp}
\affiliation{School of Engineering, Institute of Science Tokyo, 4259 Nagatsuta-cho, Midori-ku, Yokohama,
Kanagawa 226-8501, Japan}

\date{November 16, 2026}

\begin{abstract}
Boron-doped diamond (BDD) enables the integration of electrical functionality with nitrogen-vacancy (NV) centers, but the influence of junction electrostatics on NV charge-state stability remains unclear. In this study, we investigate the equilibrium electrostatics of BDD/nitrogen-doped diamond junctions using self-consistent Poisson--Fermi calculations. The results reveal two distinct electrostatic length scales in the nitrogen-doped substrate: a near-interface region of ionized donors that produces dominant band bending and determines the NV$^{0/-}$ charge-state boundary, and a much longer screening tail associated with weak donor ionization in the neutral bulk. The resulting NV$^-$ depletion depth depends strongly on the nitrogen concentration but exhibits a more moderate dependence on boron doping. 
At $N_B = 10^{21}\,\mathrm{cm^{-3}}$,
varying the interface grading scale $w$
from 0.5 to 8 nm changes the depletion
depth by approximately 22 nm,
whereas variations in the modeled bulk BDD charge response produce a shift of only 1.27 nm. These results indicate that, under the conditions studied, the nitrogen donor concentration and local interface structure primarily govern the equilibrium NV charge-state profile, rather than the bulk electronic response of BDD alone.

\end{abstract}

\maketitle

\thispagestyle{manuscriptstyle}

\section{Introduction}
\label{sec:intro}

The charge state of nitrogen-vacancy (NV) centers is critical to their operation in diamond-based quantum devices. Optical spin initialization and readout, quantum network protocols, and magnetometry generally rely on the negatively charged NV$^-$ state, which exhibits long spin coherence times \cite{Doherty2013,Balasubramanian2009,Hensen2015,Pompili2021,Rondin2014,Barry2020}. The equilibrium occupation probability
of the NV$^-$ state depends on the local
Fermi level relative to the NV$^{0/-}$
charge-transition level. Consequently, band bending near a surface or junction can alter the spatial distribution of NV charge states. Previous studies have demonstrated charge-state
control through surface termination, junction
electric fields, electrical gating, and donor
or acceptor doping
\cite{Hauf2011,Fu2010,Grotz2012,Doi2014,Schreyvogel2015,Shimizu2023,GrootBerning2014,Rose2018,Kageura2022}.
Recent measurements of boron--nitrogen co-doped
diamond films found a decreasing relative
NV$^-$ photoluminescence contribution with
increasing boron precursor ratio
\cite{Zhao2026BoronNV}.
Depth-resolved NV electrometry has also been
used to map surface band bending, whereas
charge-state measurements of shallow NV centers
have provided local probes of surface transfer
doping \cite{Broadway2018,Kageura2024}. These findings establish the importance
of local electrostatics for NV charge-state
control and motivate examining such
effects in integrated diamond devices.

Boron-doped diamond (BDD) provides a means of integrating electrical functionality into diamond devices containing NV centers. BDD offers tunable carrier densities and enables robust electrode fabrication \cite{Macpherson2015,Yang2019}, with boron concentrations exceeding $10^{21}\,\mathrm{cm^{-3}}$ achievable through doping. The boron acceptor level lies approximately 0.37 eV above the valence-band edge. As the boron concentration increases, BDD undergoes a metal--insulator transition near $4.5\times10^{20}\,\mathrm{cm^{-3}}$, with sufficiently heavily doped samples becoming superconducting at low temperatures \cite{Klein2007,Borst1996,Lagrange1998,Ekimov2004,Takano2004}. The tunable electrical properties of BDD
enable the use of buried, degenerately doped
BDD back gates and stacked BDD/NV layers
to control the NV population
\cite{Oing2021,Krecmarova2021}. BDD circuits have also been directly integrated with diamond substrates containing NV centers to serve as microwave structures \cite{Ohkuma2026Coherent}. Monolithic architectures have further demonstrated
microwave delivery and local thermometry on
substrates containing NV centers
\cite{Ohkuma2026Integrated}.
However, the equilibrium electrostatic response
of solid-state BDD/nitrogen-doped diamond junctions
remains insufficiently quantified across the
BDD metal--insulator transition.

The junction's electrostatic response
is strongly asymmetric because boron
and nitrogen have substantially different
ionization energies. Boron introduces a relatively shallow acceptor level, whereas substitutional nitrogen forms a deep donor level approximately 1.7 eV below the conduction-band edge \cite{Farrer1969,Collins2002}. Over the nitrogen concentration range considered in this study, nitrogen-doped diamond remains an insulating system characterized by dilute, deep donors, only a small fraction of which are ionized in the neutral bulk. Consequently, the equilibrium free-electron density is many orders of magnitude lower than the donor density. Near the junction, band bending can
produce a narrow region of ionized donors,
followed by a much longer, weak screening
tail as the material returns to bulk neutrality. In contrast to the weak deep-donor response
on the nitrogen-doped side, the BDD charge
response must be modeled using different
carrier statistics across the metal--insulator
transition. Conventional nondegenerate junction
theory applies below the transition, whereas
the high-density regime requires a different treatment of carrier statistics and acceptor ionization
\cite{Sze2007,ShklovskiiEfros1984,Mott1990}. These changes in the modeled charge response affect the shape of the electrostatic potential profile. The resulting profile therefore depends on the charge responses of both materials, rather than on the electronic properties of BDD alone.

For BDD/nitrogen-doped diamond junctions,
determining the equilibrium NV charge-state
boundary requires a self-consistent treatment
of the charge responses of both materials.
To address this requirement, we investigate
a one-dimensional, electrically floating
junction using self-consistent Poisson--Fermi
calculations under zero bias and in dark thermodynamic equilibrium. We examine how the nitrogen donor density, BDD charge response, and nanoscale interface grading influence the NV$^{0/-}$ charge-state boundary as BDD transitions from a nondegenerate regime to a strongly screened one. The calculations indicate that the resulting NV$^-$ depletion depth is relatively insensitive to substantial variations in the adopted high-density BDD charge response.

\section{Model and numerical method}
\label{sec:model}

\subsection{Junction geometry and electrostatic model}

The electrostatic model describes a one-dimensional junction between a nitrogen-doped diamond substrate and a boron-doped diamond (BDD) film. The substrate occupies $z < 0$, whereas the BDD film occupies $0 < z < t_{\mathrm{BDD}}$, with $t_{\mathrm{BDD}} = 1\,\mu\mathrm{m}$.
Our calculation proceeds in two stages.
First, we determine the electrostatic potential and the global Fermi level
self-consistently using Poisson's equation, carrier statistics, dopant
ionization, and global charge neutrality.
We then evaluate the equilibrium NV charge-state occupation from the
resulting band profile.

The electrostatic potential $\psi(z)$ satisfies Poisson's equation:
\begin{equation}
\frac{d}{dz}\left(\varepsilon\frac{d\psi}{dz}\right)=-\rho(z),
\label{eq:poisson}
\end{equation}
where the local charge density is given by
\begin{equation}
\rho=q\left(p-n+N_D^+-N_A^-\right)+\rho_{\mathrm{int}}(z).
\label{eq:rho}
\end{equation}
Here, $n$ and $p$ are the free-electron and hole densities, respectively, and $N_D^{+}$ and $N_{A}^{-}$ are the ionized nitrogen donor and boron acceptor densities, respectively. The term $\rho_{\mathrm{int}}(z)$ represents an optional fixed interface sheet charge $Q_{\mathrm{int}}$.

The dopant and interface-charge terms in Eq.~\eqref{eq:rho} are specified through prescribed spatial profiles. We model the active dopant concentrations as varying smoothly across the interface according to
\begin{equation}
\begin{aligned}
N_D(z)&=N_{D0}s_D(z),&
N_A(z)&=N_{B0}[1-s_D(z)],\\
s_D(z)&=\frac{1-\tanh(z/w)}{2}.&&
\end{aligned}
\label{eq:profiles}
\end{equation}
Here, $N_{B0}$ and $N_{D0}$ denote the bulk active boron and substitutional nitrogen concentrations, respectively. The parameter $w$ controls the interface width, with a 10--90\% transition width of $2.20w$. This hyperbolic tangent profile is phenomenological and does not represent a microscopic model of the dopant distributions. We adopt $w = 8\,\mathrm{nm}$ as a
representative reference value for
the sensitivity analysis. Throughout this study, boron concentrations are reported as $N_B=N_{B0}$ in $\mathrm{cm^{-3}}$, whereas nitrogen concentrations are reported as $[\mathrm N_s]$ in ppm, with 1 ppm corresponding to $N_{D0}=1.76\times10^{17}\,\mathrm{cm^{-3}}$. Relating these active concentrations to nominal growth concentrations or secondary-ion mass spectrometry (SIMS) measurements requires information on dopant activation and compensation. The optional fixed interface charge is represented by a normalized Gaussian profile with a standard deviation of 2 nm, as specified in Sec.~I of the Supplemental Material \cite{SupplementalMaterial}.

The electrostatic problem is constrained by the electrically floating nature of the junction, with both outer surfaces electrically neutral. Its equilibrium Fermi level $E_F$ is set by global charge neutrality rather than by an external contact. This Fermi level is spatially uniform and is determined self-consistently with the electrostatic potential. We impose a zero electric field at the substrate boundary and set $\psi = 0$ at the BDD surface to fix the potential gauge. The electric field at the BDD surface is not imposed independently; instead, it follows from Poisson's equation and global charge neutrality. The consistency of this boundary field is verified after convergence.

\subsection{Carrier statistics and doping regimes}

To solve this electrostatic problem, we evaluate the free-carrier and ionized-dopant terms in Eq.~\eqref{eq:rho} using models for nondegenerate and high-density BDD regimes. The band energies are expressed in electron volts. When $\psi$ is expressed in volts, the conduction-band edge is given by $E_C(z) = E_{C0} - \psi(z)$, corresponding to an electrostatic energy shift of $-q\psi$ in joules. The valence-band edge is $E_V = E_C - E_{g}$. Below the BDD metal--insulator transition, the free-carrier populations follow nondegenerate statistics:
\begin{equation}
\begin{aligned}
n&=N_C\exp[(E_F-E_C)/k_BT],\\
p&=N_V\exp[(E_V-E_F)/k_BT],
\end{aligned}
\label{eq:carriers}
\end{equation}
whereas incomplete dopant ionization is described by
\begin{equation}
\begin{aligned}
N_D^+&=\frac{N_D(z)}{1+g_D\exp[(E_F-E_D)/k_BT]},\\
N_A^-&=\frac{N_A(z)}{1+g_A\exp[(E_A-E_F)/k_BT]}.
\end{aligned}
\label{eq:dopants}
\end{equation}
The adopted parameters are
$E_C-E_D=1.7\,\mathrm{eV}$,
$E_A-E_V=0.37\,\mathrm{eV}$,
$g_D=2$, and $g_A=4$
(Table~\ref{tab:parameters}). Using a hydrogenic acceptor Bohr radius of $a_B^*=0.341\,\mathrm{nm}$, the Mott criterion~\cite{Mott1990} $N_{B,\mathrm{Mott}}^{1/3}a_B^*=0.26$ gives $N_{B,\mathrm{Mott}}=4.4\times10^{20}\,\mathrm{cm^{-3}}$, which we use as a reference boundary for the isolated-acceptor model.

For the adopted high-density treatment, the BDD response is instead described using fully ionized acceptors and finite-temperature Fermi--Dirac statistics:
\begin{equation}
\begin{aligned}
p&=N_VF_{1/2}\!\left(\frac{E_V-E_F}{k_BT}\right),\\
n&=N_CF_{1/2}\!\left(\frac{E_F-E_C}{k_BT}\right).
\end{aligned}
\label{eq:fd}
\end{equation}
Here, $F_{1/2}$ is normalized to approach $\exp(\eta)$ in the Boltzmann limit. We use nondegenerate and high-density treatments to examine the electrostatic response below and above the metal--insulator transition, respectively, without modeling the transition's critical microscopic behavior. The high-density treatment is not assumed to reproduce the microscopic density of states of heavily doped BDD. Accordingly, quantitative high-density results are restricted to $N_B \leq 10^{21}\,\mathrm{cm^{-3}}$; calculations at $10^{22}\,\mathrm{cm^{-3}}$ are presented only as parabolic-band extrapolations.

\subsection{Equilibrium NV charge occupation}

The calculated band profile determines the equilibrium occupation probability of the negatively charged NV state. We express the NV$^-$ occupation probability in terms of the local NV$^{0/-}$ charge-transition level and the Fermi level as

\begin{equation}
P(\mathrm{NV}^-;z)=
\left[1+\exp\!\left(\frac{E_{\mathrm{NV}^{0/-}}(z)-E_F}{k_BT}\right)\right]^{-1},
\label{eq:pnv}
\end{equation}
where $E_{\mathrm{NV}^{0/-}}(z) = E_C(z) - \Delta E_{\mathrm{NV}}$. We use $\Delta E_{\mathrm{NV}} =
2.7\,\mathrm{eV}$ and vary it over
2.6--2.8 eV to account for the
reported spread in the NV
charge-transition level
\cite{Hauf2011,Aslam2013,Deak2014}.

Equation~\eqref{eq:pnv} treats the specified transition level as an effective finite-temperature free-energy level with a statistical prefactor of unity. For a bare transition energy $E_t$ and explicit state degeneracies $g_0$ and $g_-$, the denominator becomes $1+(g_0/g_-)\exp[(E_t-E_F)/(k_BT)]$. This expression is equivalent to using an effective transition level of $E_t+k_BT\ln(g_0/g_-)$. We account for this convention within the specified transition-level range rather than introducing it as an additional independent parameter.

The NV$^-$ depletion depth $d_{\mathrm{NV^-}}$ is defined as the distance into the substrate at which $P\left( \mathrm{NV}^- \right) = 1/2$. This depth identifies the equilibrium NV$^{0/-}$ charge-state boundary relevant to quantum-sensing applications. The NV$^{0/-}$ transition alone does not establish whether NV centers closer to the interface are neutral (NV$^0$) or positively charged (NV$^+$).

Independently of this charge-state
assignment, we treat NV centers as
dilute probes and therefore omit their
charge contributions from Eq.~\eqref{eq:rho}.
This approximation requires the NV sheet density
to be much smaller than the ionized-donor sheet
density in the space-charge region.
Assuming substantial donor ionization over a
characteristic width $d_{\mathrm{dep}}$, an
order-of-magnitude criterion is
\begin{equation}
N_{\mathrm{NV}} t_{\mathrm{NV}} \ll N_{D0} d_{\mathrm{dep}},
\label{eq:nv_dilute}
\end{equation}
where $N_{\mathrm{NV}}$ and $t_{\mathrm{NV}}$
denote the NV volume density and the thickness
of the NV-containing layer, respectively.
For a nitrogen concentration of 1 ppm and
$d_{\mathrm{dep}} = 100\,\mathrm{nm}$,
a 10\% criterion gives
$N_{\mathrm{NV}}t_{\mathrm{NV}} <
1.8 \times 10^{11}\,\mathrm{cm^{-2}}$,
corresponding to approximately 0.1 ppm in a
uniform 100-nm-thick NV layer.
If this dilute-probe condition is not satisfied,
the NV charge must be included self-consistently
in the electrostatic model.

\begin{table}[t]
\caption{Parameters used in the electrostatic calculations.}
\label{tab:parameters}
\begin{ruledtabular}
\begin{tabular}{lll}
Quantity & Symbol & Value \\
\hline
Band gap & $E_g$ & 5.47 eV \\
Relative permittivity & $\varepsilon_r$ & 5.7 \\
Temperature & $T$ & 300 K \\
Effective band DOS & $N_C,N_V$ & $10^{19}\,\mathrm{cm^{-3}}$ \\
Nitrogen donor level & $E_C-E_D$ & 1.7 eV \\
Boron acceptor level & $E_A-E_V$ & 0.37 eV \\
Dopant degeneracies & $g_D,g_A$ & 2, 4 \\
NV$^{0/-}$ level & $E_C-E_{\mathrm{NV}^{0/-}}$ & $2.7\pm0.1$ eV \\
BDD thickness & $t_{\mathrm{BDD}}$ & 1 $\mu$m \\
Interface grading scale & $w$ & 8 nm \\
Interface charge range & $|Q_{\mathrm{int}}|/q$ & $\leq10^{12}\,\mathrm{cm^{-2}}$ \\
\end{tabular}
\end{ruledtabular}
\end{table}

\subsection{Numerical solution}

The coupled Poisson and charge-neutrality equations are solved using finite-volume discretization on a nonuniform mesh. The default mesh spacing is 2 nm within 2 $\mu$m of the interface and throughout the BDD film, increasing to 20 nm and then to 1 $\mu$m in the remote substrate. For the small-signal screening analysis,
the mesh spacing near the interface and
within the BDD film is reduced to 0.5 nm,
with additional refinement checks down
to 0.25 nm.
Cells intersecting the interface are divided at $z = 0$, as specified in Sec.~VI of the Supplemental Material \cite{SupplementalMaterial}. The high-density BDD calculation uses a 0.05-nm mesh over the first 20 nm of the BDD region to resolve the calculated subnanometer screening length. Resolving this length scale numerically is essential for solution accuracy, although it does not establish the physical validity of a continuum dielectric model at the atomic scale.

The electrostatic potential and Fermi level are then determined using a damped Newton method with continuation of the carrier response. Each solution is checked for global charge neutrality, Gauss's law closure, local finite-volume residuals, and changes between Newton iterations. To ensure that the domain size does not affect the results, we explicitly verify the convergence of local observables with respect to the substrate length. The production-domain criteria
$|\rho|/(qN_{D0}) < 10^{-4}$ and
$|E| < 10^{3}\,\mathrm{V\,m^{-1}}$
are used only to select an initial trial
domain, not as convergence criteria.
Convergence of $d_{\mathrm{NV^-}}$, $E_F$, and the interface potential is verified through fixed-domain calculations extending to 10 mm for the nondegenerate model and 3 mm for the high-density model, as documented in Table~S1 of the Supplemental Material \cite{SupplementalMaterial}.

\section{Results}
\label{sec:results}

\subsection{Two electrostatic length scales and NV charge occupation}

\begin{figure}[!t]
\includegraphics[width=\columnwidth]{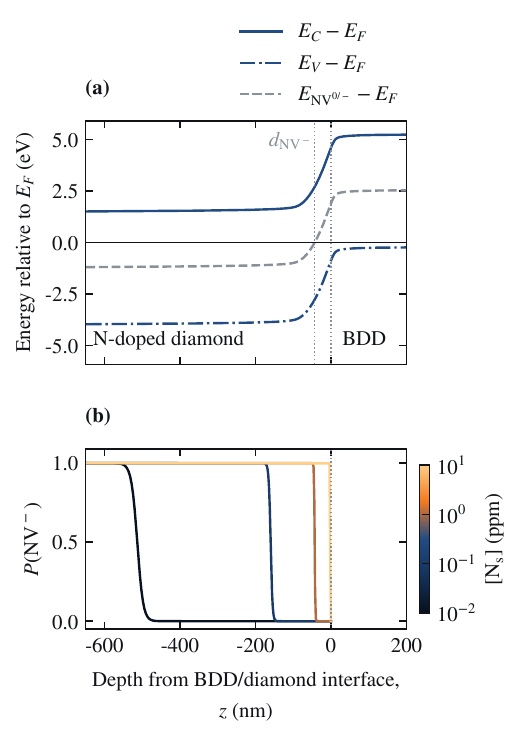}
\caption{(a) Band edges and NV$^{0/-}$ transition level for $N_B=10^{18}\,\mathrm{cm^{-3}}$, $[\mathrm N_s]=1$ ppm, and $Q_{\mathrm{int}}=0$ at 300 K. Energies are measured relative to the global Fermi level. The interface is located at $z=0$, with nitrogen-doped diamond at $z<0$. (b) Equilibrium NV$^-$ occupation probability for $[\mathrm N_s]=0.01$--10 ppm at a fixed $N_B=10^{18}\,\mathrm{cm^{-3}}$. The NV$^{0/-}$ crossing identifies the loss of NV$^-$ occupation but does not uniquely establish that the depleted centers are in the NV$^0$ state.}
\label{fig:profile}
\end{figure}

The calculated equilibrium band profile places the
NV$^{0/-}$ charge-state boundary within the
near-interface region of the nitrogen-doped substrate. For a representative junction with $N_B = 10^{18}\,\mathrm{cm^{-3}}$, $[\mathrm N_s] = 1$ ppm ($N_{D0} = 1.76 \times 10^{17}\,\mathrm{cm^{-3}}$), and $Q_{\mathrm{int}} = 0$, the isolated bulk Fermi levels differ by approximately 4.3 eV, resulting in band bending predominantly on the nitrogen-doped side (Fig.~\ref{fig:profile}(a)). The NV$^{0/-}$ transition level crosses the global Fermi level at $d_{\mathrm{NV^-}} = 43$ nm, where the calculated NV$^-$ occupation probability changes rapidly with depth [Fig.~\ref{fig:profile}(b)].

The near-interface band bending that determines the
NV$^{0/-}$ charge-state boundary occurs within a
region of ionized nitrogen donors extending
approximately 100 nm into the substrate.
Beyond this region, the substrate approaches bulk
neutrality through a much weaker electrostatic tail. Linearization of the charge response yields an asymptotic screening length of $\lambda_{\infty} \sim 2.87$ cm at a nitrogen concentration of 1 ppm, approximately five orders of magnitude larger than the near-interface length scale (Supplemental Material, Sec.~III, Eq.~(S10) \cite{SupplementalMaterial}). This unusually long screening length results from the freeze-out of deep donors. Although the total nitrogen concentration is substantial, the equilibrium differential charge response is governed by the very small populations of free electrons and ionized donors. 
In the substrate-length sweep,
the local NV$^{0/-}$ charge-state boundary
converges as the substrate length increases,
whereas the total potential drop between the two
outer surfaces does not converge over the tested
substrate lengths (Supplemental Material,
Fig.~S1(a,d) \cite{SupplementalMaterial}).
This difference in convergence behavior reflects the separation between
the near-interface and bulk screening responses. The NV charge-state
boundary is primarily determined by the localized band bending associated
with donor ionization near the junction. The total potential drop includes
the integral of the weak electric field in the long-range screening tail
and therefore remains sensitive to the position of the outer substrate
boundary.

\subsection{Nitrogen concentration controls the NV$^-$ depletion depth}

\begin{figure}[tbp]
\includegraphics[width=\columnwidth]{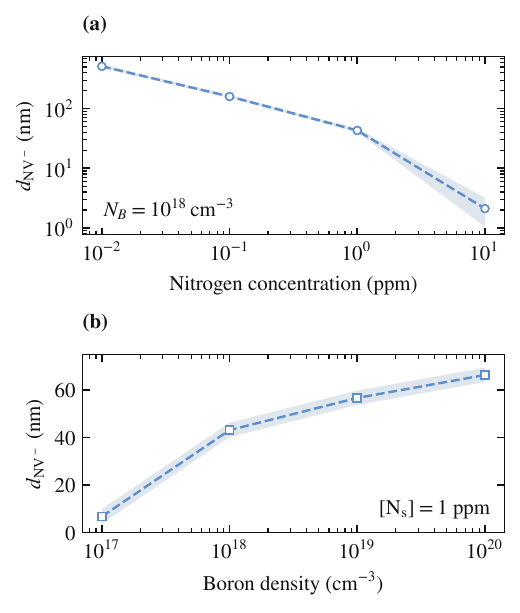}
\caption{Calculated NV$^-$ depletion depth (a) as a function of nitrogen concentration at $N_B=10^{18}\,\mathrm{cm^{-3}}$ and (b) as a function of boron concentration at $[\mathrm N_s]=1$ ppm. Shading indicates the range obtained for $E_C-E_{\mathrm{NV}^{0/-}}=2.6$--$2.8$ eV. The substantially stronger dependence on
nitrogen concentration reflects the strongly
asymmetric, effectively one-sided electrostatic
response of the junction.}
\label{fig:depth}
\end{figure}

Over the ranges examined, the calculated NV$^-$ depletion depth depends more strongly on the nitrogen concentration than on the boron concentration (Fig.~\ref{fig:depth} and Table~\ref{tab:depth}). At a fixed $N_B = 10^{18}\,\mathrm{cm^{-3}}$, increasing $[\mathrm N_s]$ from 0.01 to 10 ppm reduces $d_{\mathrm{NV^-}}$ from 513 nm to approximately 2 nm, a value comparable to the production mesh spacing [Fig.~\ref{fig:depth}(a)]. At 100 ppm, the NV$^{0/-}$ transition level does not cross $E_F$, and no NV$^-$ depletion layer forms in the calculation. In contrast, at a fixed $[\mathrm N_s] = 1$ ppm, increasing $N_B$ from $10^{17}$ to $10^{20}\,\mathrm{cm^{-3}}$ increases $d_{\mathrm{NV^-}}$ from 6.74 to approximately 66 nm [Fig.~\ref{fig:depth}(b)]. Once the junction becomes strongly one-sided,
further increases in the boron concentration
primarily shift the BDD chemical potential
and produce smaller changes in the substrate
depletion width.

The calculated depletion depth of
6.74 nm at $N_B = 10^{17}\,\mathrm{cm^{-3}}$
requires sufficient spatial resolution.
We therefore assess mesh convergence
for this case.  
At a fixed $L_{\mathrm{sub}} = 100$~$\mu$m,
reducing the near-interface mesh spacing
from 2.0 to 1.0 and 0.5 nm yields
depletion depths of 6.7395, 6.7369,
and 6.7366 nm, respectively.
The difference between the coarsest and finest mesh results is 0.0030 nm (0.044\%), confirming mesh convergence for this case (Supplemental Material, Table~S2 \cite{SupplementalMaterial}).

\begin{table}[t]
\caption{NV$^-$ depletion depth for $E_C-E_{\mathrm{NV}^{0/-}}=2.7$ eV. The Range column reports the values obtained for transition energies of 2.6--2.8~eV. A dash indicates that the NV$^{0/-}$ transition level does not cross $E_F$.}
\label{tab:depth}
\begin{ruledtabular}
\begin{tabular}{llll}
$N_B$ (cm$^{-3}$) & $[\mathrm{N}_{\mathrm{s}}]$ (ppm) & $d_{\mathrm{NV^-}}$ (nm) & Range (nm) \\
\hline
$10^{16}$ & 1 & -- & -- \\
$10^{17}$ & 1 & 6.74 & 4--10 \\
$10^{18}$ & 1 & 43 & 40--46 \\
$10^{19}$ & 1 & 57 & 54--60 \\
$10^{20}$ & 1 & 66 & 63--69 \\
$10^{18}$ & 0.01 & 513 & 483--544 \\
$10^{18}$ & 0.1 & 160 & 150--169 \\
$10^{18}$ & 10 & $\sim2$ & 1.05--3.22 \\
$10^{18}$ & 100 & -- & -- \\
\end{tabular}
\end{ruledtabular}
\end{table}

\subsection{BDD screening across the metal--insulator transition}

\begin{figure*}[!t]
\includegraphics[width=\textwidth]{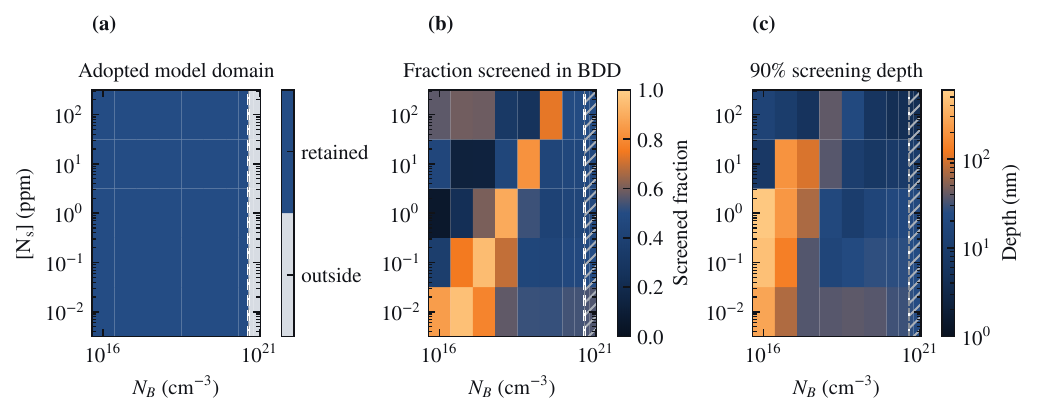}
\caption{Small-signal screening in the nondegenerate isolated-acceptor model: (a) parameter domain satisfying the numerical acceptance criteria, carrier-statistics checks, and $N_B<N_{B,\mathrm{Mott}}$; (b) signed fraction of the applied charge screened in the region $z>0$; and (c) depth containing 90\% of the absolute induced material charge in that region. The dashed line indicates the reference Mott density, $N_{B,\mathrm{Mott}} = 4.4\times10^{20}\,\mathrm{cm}^{-3}$. Hatched regions in (b) and (c) indicate extrapolations beyond the validity range of the isolated-acceptor model.}
\label{fig:screening}
\end{figure*}

The redistribution of induced charge across the junction
is characterized within the nondegenerate isolated-acceptor
model by the signed screening fraction and the
90\% BDD screening depth (Fig.~\ref{fig:screening}).
The signed screening fractions are calculated by integrating the induced charge over the nitrogen-doped ($z < 0$) and BDD ($z > 0$) regions and normalizing by the negative of the applied sheet charge. The 90\% BDD screening depth is defined as the depth containing 90\% of the absolute induced material charge in the region $z > 0$, excluding the imposed charge.

In the nondegenerate model, substantial
screening in the BDD region shifts toward
higher boron concentrations as the nitrogen
concentration increases
[Fig.~\ref{fig:screening}(b)].
The calculated 90\% BDD screening depth
also depends strongly on both boron and
nitrogen concentrations
[Fig.~\ref{fig:screening}(c)]. At $N_B = 10^{16}\,\mathrm{cm^{-3}}$, the 90\% screening depth ranges from 7.2 to 516.2 nm as $[\mathrm N_s]$ increases from 0.01 to 100 ppm. For the sampled points with $10^{18} \lesssim N_B \lesssim 2 \times 10^{20}\,\mathrm{cm^{-3}}$, the screening depth ranges from 4.2 to 38.5 nm over the same nitrogen concentration range, including values of 30.4--38.4 nm at 0.01 ppm. These screening depths are characteristic of semiconductor charge redistribution and do not, by themselves, imply metallic transport.

These nondegenerate screening trends must, however, be interpreted within the validity range of the isolated-acceptor model. All 55 sampled points passed the numerical carrier-statistics checks, including 15 points at or above the reference Mott density, $N_{B,\mathrm{Mott}}$. Passing these checks does not establish
the physical validity of the isolated-acceptor
model at high doping concentrations. We therefore retain points above $N_{B,\mathrm{Mott}}$ only to illustrate the approach to the model's validity limit, not for quantitative predictions. The model is considered quantitatively applicable only for $N_B < N_{B,\mathrm{Mott}}$; the hatched regions in Fig.~\ref{fig:screening}(b) and (c) indicate extrapolations beyond this range. To evaluate the high-density response
beyond the isolated-acceptor model's
adopted validity range, we instead use
the Fermi--Dirac treatment in
Eq.~\eqref{eq:fd}.
This treatment does not attempt to describe
the critical microscopic behavior of the
metal--insulator transition.

We next verify the numerical implementation of the adopted high-density model against its analytical screening limit. The Fermi--Dirac carrier statistics recover the Boltzmann limit under the same acceptor-ionization assumptions. For homogeneous BDD, the calculated screening length agrees with the analytical finite-temperature Thomas--Fermi screening length. At $N_B = 10^{21}\,\mathrm{cm^{-3}}$ and $w = 0.5$~nm, the fitted and analytical screening lengths are 0.383 and 0.376~nm, respectively. The corresponding 90\% screening depths are 0.882 and 0.866 nm, respectively. Refining the BDD mesh spacing from 0.10 to 0.05 nm does not change these values at the reported precision. This agreement verifies the numerical implementation of the adopted continuum model but does not establish its microscopic validity at subnanometer scales, where atomic-scale band structure and dielectric response are not represented.

\begin{figure}[tb]
\includegraphics[width=\columnwidth]{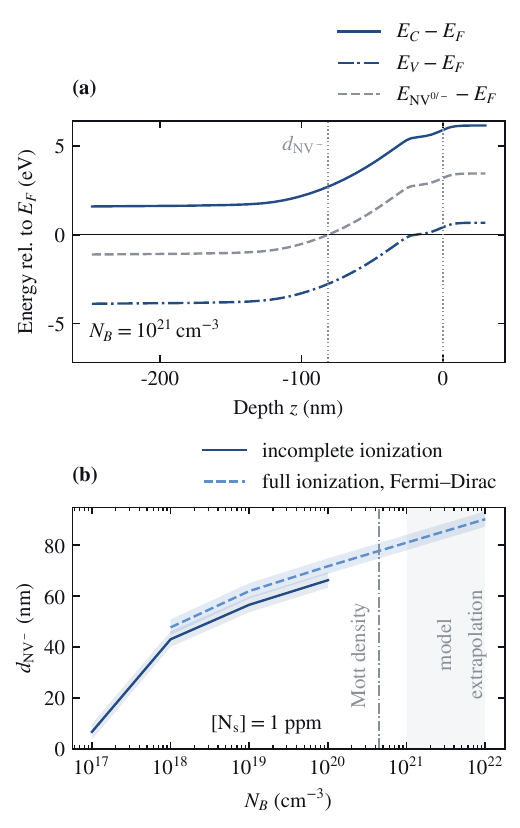}
\caption{(a) Band diagram calculated using the adopted high-density model for $N_B=10^{21}\,\mathrm{cm^{-3}}$ and $[\mathrm N_s]=1$ ppm. (b) NV$^-$ depletion depth predicted by the incomplete-ionization model and the degenerate, fully ionized model. Shading indicates the range associated with the NV transition level, and the vertical line marks the reference Mott density. Results above $10^{21}\,\mathrm{cm^{-3}}$ are parabolic-band extrapolations. The $10^{22}\,\mathrm{cm^{-3}}$ point is included only to illustrate the asymptotic behavior of the adopted model and is not used as a quantitative prediction.}
\label{fig:highdensity}
\end{figure}

Near the metal--insulator transition,
the calculated NV$^-$ depletion depth
depends on the adopted treatment of
the BDD charge response
[Fig.~\ref{fig:highdensity}(b)].
Within the adopted high-density treatment,
the depletion depth changes only modestly
as the boron concentration increases
above the reference Mott density.
Specifically, the model yields $d_{\mathrm{NV^-}} = 78$ nm at $N_{B,\mathrm{Mott}}$ and 81 nm at $N_B = 10^{21}\,\mathrm{cm^{-3}}$. At the latter concentration, the calculated Fermi level lies 0.67 eV below the valence-band edge. Extrapolating the same parabolic-band model to $N_B = 10^{22}\,\mathrm{cm^{-3}}$ yields $E_V - E_F \simeq 3.1$ eV. However, this concentration lies outside the quantitative range of the present study. 
Thus, within the adopted high-density model and its stated quantitative range, increasing the boron concentration produces only a modest change in the calculated $\mathrm{NV^-}$ depletion depth.

\subsection{Interface sensitivity in the strongly screened regime}

\begin{figure*}[!t]
\includegraphics[width=\textwidth]{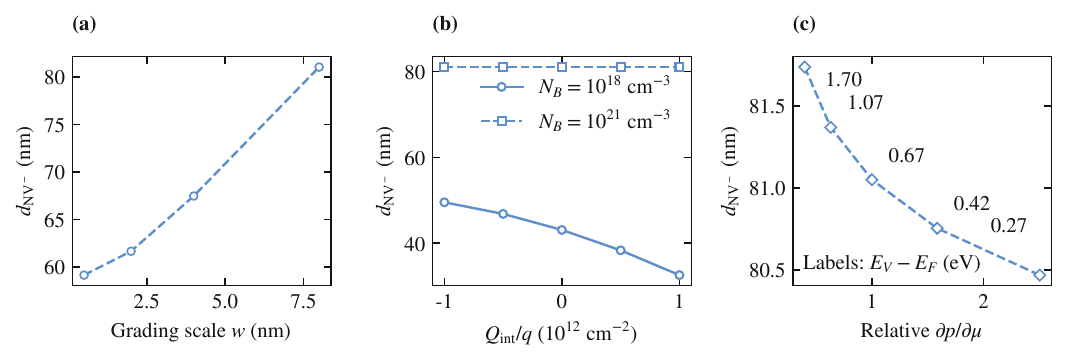}
\caption{Sensitivity of $d_{\mathrm{NV^-}}$ at $[\mathrm N_s]=1$ ppm: (a) interface grading scale at $N_B=10^{21}\,\mathrm{cm^{-3}}$; (b) fixed interface charge at $N_B=10^{18}$ and $10^{21}\,\mathrm{cm^{-3}}$, with $w=8$ nm; and (c) high-density BDD charge response varied through the effective density of states. Labels in (c) indicate the corresponding bulk values of $E_V-E_F$ in electron volts. The NV$^-$ depletion depth is substantially more sensitive to nanoscale interface grading than to large variations in the modeled high-density BDD response.}
\label{fig:sensitivity}
\end{figure*}

The limited variation in the calculated
NV$^-$ depletion depth with boron concentration
in the adopted high-density regime motivates
a comparison of its sensitivity to interface
grading, fixed interface charge, and the
modeled bulk BDD charge response
(Fig.~\ref{fig:sensitivity}).
At $N_B = 10^{21}\,\mathrm{cm^{-3}}$, varying the interface grading scale $w$ from 0.5 to 8 nm increases $d_{\mathrm{NV^-}}$ from 59.15 to 81.05 nm [Fig.~\ref{fig:sensitivity}(a)]. The intermediate values are 61.66 nm at $w = 2$ nm and 67.47 nm at $w = 4$ nm. The resulting variation of approximately 22 nm substantially exceeds the approximately 3-nm change associated with the $\pm0.1$ eV range of the NV charge-transition level.
Changing the grading scale $w$ by a few nanometers redistributes ionized
dopants near the junction. In the high-density model, increasing $w$
extends the distribution of fully ionized boron acceptors farther into
the nitrogen-doped substrate. This redistribution of space charge changes
the self-consistent electrostatic potential and shifts the crossing of the
local NV transition level with $E_F$. The displacement depends on both
the change in the energy difference $E_{\mathrm{NV}^{0/-}}(z)-E_F$
and its slope near the crossing, so it can exceed the change in $w$.

Interface charge provides a second
interfacial perturbation, but its effect
depends on the BDD screening regime. At $N_B = 10^{18}\,\mathrm{cm^{-3}}$, an interface charge of $Q_{\mathrm{int}}/q = \pm 10^{12}\,\mathrm{cm^{-2}}$ changes the depletion depth from 43.09 nm to values ranging from 32.57 to 49.53 nm [Fig.~\ref{fig:sensitivity}(b)]. At $N_B = 10^{21}\,\mathrm{cm^{-3}}$, the same centered Gaussian charge distribution changes the depletion depth by less than $10^{-3}$ nm because the charge is screened locally. This negligible change applies specifically to the assumed one-dimensional geometry and centered 2-nm Gaussian charge distribution; an asymmetric interface-charge distribution may produce a different response.

The calculated depletion depth is also
relatively insensitive to variations in
the modeled high-density BDD charge
response.
To evaluate this response, we scale
$N_{C} = N_{V}$ from 0.25 to 4 times
the values listed in
Table~\ref{tab:parameters}.
This variation changes the calculated bulk $E_V - E_F$ from 1.697 to 0.265 eV and changes the differential charge response by a factor of 6.3. Nevertheless, $d_{\mathrm{NV^-}}$ changes only from 81.74 to 80.47 nm [Fig.~\ref{fig:sensitivity}(c)], a total variation of 1.27 nm. Under the conditions examined, these results indicate that nanoscale interface structure has a greater influence on the equilibrium NV charge-state boundary than variations in the modeled bulk BDD electronic response.

\section{Discussion}
\label{sec:discussion}

The electrostatic influence of BDD on
NV charge-state stability depends on its
charge response relative to that of the
nitrogen-doped substrate. The nitrogen
donor density establishes the principal
space-charge scale, whereas the influence
of the bulk BDD charge response diminishes
once its screening length becomes
sufficiently short. In this regime, the nanoscale interface structure, particularly dopant grading, has a greater influence on the equilibrium NV$^{0/-}$ charge-state boundary than variations in the modeled bulk BDD electronic response. The coupling of interfacial electrochemical potentials to embedded defects has also been examined theoretically in diamond–electrolyte systems \cite{Kimura2026Universal}. The predicted nitrogen-concentration dependence
is qualitatively consistent with observations
in surface-transfer-doped diamond, where shallow
NV centers probe nitrogen-dependent band bending
\cite{Kageura2024}.
Nevertheless, the present calculations establish
a dark-equilibrium baseline rather than directly
describing optically driven NV charge-state measurements.

These modeled charge-state trends provide
a basis for assessing NV placement in
BDD-integrated diamond structures.
For a substrate containing 1 ppm nitrogen and $N_B = 10^{21}\,\mathrm{cm^{-3}}$, the interface-grading profiles examined here yield $d_{\mathrm{NV^-}} = 59$--81 nm. Varying the NV charge-transition level over the adopted range introduces an additional change of approximately 3 nm. An NV layer located at a depth of 150 nm would therefore lie beyond the depletion region under these specific equilibrium conditions. However, this depth is neither an intrinsic material property nor a universal device-design threshold. Quantitative predictions for a particular
heterostructure require experimentally constrained
active dopant concentrations, compensation levels,
and nanoscale BDD/diamond composition profiles,
together with appropriate interface-charge
parameters and electrical boundary conditions.

The equilibrium charge-state boundary used
in this placement estimate should not be
interpreted as an optical switching boundary. Our calculations describe dark thermodynamic equilibrium at 300 K, whereas optically driven NV populations depend on charge-transfer kinetics. Frozen dopant or trap configurations may retain a dependence on their preparation history, and optical excitation can establish a steady state governed by photoionization, recombination, tunneling, and trap release \cite{Aslam2013}. Optical charge transfer has been shown to stabilize shallow NV$^-$ centers in lightly boron-doped diamond \cite{Kageura2022}, whereas charge injection can coexist with photoionization in electrically operated hydrogen-terminated diamond \cite{Broadway2018}. Predicting the NV charge-state distribution under these conditions requires a kinetic model beyond the equilibrium treatment employed here. Even under dark equilibrium conditions, the continuum occupation model does not resolve local donor--NV correlations, which may influence the charge states of individual defects in insulating diamond \cite{Collins2002}.

The present calculations assume an electrically floating, globally neutral, one-dimensional stack. In an operating device, electrical contact with the BDD layer can constrain its electrochemical potential, whereas an applied bias or grounding can alter the junction potential and NV charge-state profile. The one-dimensional approximation is also limited to the interior of a laterally extended film. Near electrode edges or in patterned circuits, lateral fringing fields and crosstalk require two- or three-dimensional calculations with explicit contacts and device geometry.

In addition to these geometric and electrical limitations, the adopted high-density BDD model is phenomenological and does not resolve the microscopic electronic structure. Heavily boron-doped diamond can exhibit
anisotropic valence-band structure and
associated non-spherical Fermi surfaces
that are not explicitly resolved by the
single effective-density-of-states
description adopted here~\cite{Kawano2010}. The limited sensitivity of the calculated NV$^-$ depletion depth to variations in the modeled bulk BDD charge response supports the robustness of the identified electrostatic hierarchy over the range examined, but does not establish the microscopic accuracy of the constitutive model. In particular, the parabolic-band approximation, complete acceptor ionization, and continuum dielectric response have not been validated for arbitrary boron concentrations. Although the numerical implementation is consistent with Eq.~\eqref{eq:fd}, the sensitivity analysis does not justify quantitatively extrapolating the results to $N_B = 10^{22}\,\mathrm{cm^{-3}}$.

\section{Conclusion}
\label{sec:conclusion}

In this study, we investigated the equilibrium electrostatics of electrically floating BDD/nitrogen-doped diamond junctions using self-consistent Poisson–Fermi calculations. The results reveal that the nitrogen-doped substrate exhibits two distinct electrostatic length scales: a near-interface region of ionized donors that governs the NV$^{0/-}$ charge-state boundary and a much longer, weak screening tail associated with deep-donor freeze-out. The resulting NV$^-$ depletion depth depends strongly on the nitrogen concentration, whereas its dependence on boron concentration weakens in the strongly screened regime. At high boron concentrations, nanoscale interface grading has a greater influence on the depletion depth than variations in the modeled bulk BDD charge response.

These findings apply to one-dimensional, electrically floating junctions under zero-bias, dark thermodynamic equilibrium, with dilute NV centers and active boron concentrations up to $10^{21}\,\mathrm{cm^{-3}}$. Within the adopted equilibrium model, a
strongly screened BDD layer does not
necessarily make its bulk charge response
the dominant factor governing nearby
NV charge states. Once screening in that layer becomes sufficiently
short-ranged, the charge environment can instead
be governed by deep dopants and nanoscale
interface structure in the surrounding material.

\begin{acknowledgments}

The authors thank Taisuke Kageura for valuable discussions and insightful comments on the manuscript. This work was supported by JST K Program Grant Number JPMJKP24F3, JST FOREST Program Grant Number JPMJFR2547, and JSPS KAKENHI Grant Number JP26K22715. E.K. was supported by JST SPRING, Grant Number JPMJSP2180.

We acknowledge the use of ChatGPT
(OpenAI, GPT-6) to assist in refining
the language, clarity, and logical organization
of this manuscript. The tool was used to suggest
revisions while preserving the scientific meaning,
and the authors reviewed the revisions for
scientific accuracy and consistency with the
underlying calculations.

\end{acknowledgments}

\section*{Author Contributions}

M.K. extended the theoretical framework, developed and validated the numerical implementation, performed the calculations and analyses, and prepared the manuscript.
H.C. conceived the original BDD junction model, developed the initial self-consistent electrostatic framework, and established the connection between junction band bending and NV charge-state conversion.
E.M.M. contributed to theoretical validation, formal analysis, physical interpretation, and model refinement.
E.K. contributed to project development and provided input on the experimental and physical interpretations.
R.M. and Y.T. contributed expertise in boron-doped diamond materials and helped interpret the model's physical implications.
K.A. conceived and supervised the project and contributed to the theoretical framework, interpretation, and manuscript preparation.
All authors discussed the results and reviewed the manuscript.

\section*{Data Availability}

The numerical data and scripts used to reproduce the figures and convergence tests are available from Zenodo~\cite{Koga2026Data}.

\section*{Conflict of Interest}

The authors declare no conflict of interest.

\bibliographystyle{aipnum4-2}
\makeatletter
\immediate\write\@auxout{\string\citation{aip41Control}}
\makeatother
\bibliography{references}

@CONTROL{aip41Control, author = {48}, editor = {1}, pages = {1}, title = {0}, year = {1}}

@article{Doherty2013,
  author = {Doherty, Marcus W. and Manson, Neil B. and Delaney, Paul and Jelezko, Fedor and Wrachtrup, J{\"o}rg and Hollenberg, Lloyd C. L.},
  title = {The nitrogen-vacancy colour centre in diamond},
  journal = {Phys. Rep.},
  volume = {528},
  pages = {1--45},
  year = {2013},
  doi = {10.1016/j.physrep.2013.02.001},
}

@article{Balasubramanian2009,
  author = {Balasubramanian, Gopalakrishnan and Neumann, Philipp and Twitchen, Daniel and Markham, Matthew and Kolesov, Roman and Mizuochi, Norikazu and Isoya, Junichi and Achard, Jocelyn and Beck, Johannes and Tissler, Julia and Jacques, Vincent and Hemmer, Philip R. and Jelezko, Fedor and Wrachtrup, J{\"o}rg},
  title = {Ultralong spin coherence time in isotopically engineered diamond},
  journal = {Nat. Mater.},
  volume = {8},
  pages = {383--387},
  year = {2009},
  doi = {10.1038/nmat2420},
}

@article{Hensen2015,
  author = {Hensen, B. and Bernien, H. and Dr{\'e}au, A. E. and Reiserer, A. and Kalb, N. and Blok, M. S. and Ruitenberg, J. and Vermeulen, R. F. L. and Schouten, R. N. and Abell{\'a}n, C. and Amaya, W. and Pruneri, V. and Mitchell, M. W. and Markham, M. and Twitchen, D. J. and Elkouss, D. and Wehner, S. and Taminiau, T. H. and Hanson, R.},
  title = {Loophole-free {Bell} inequality violation using electron spins separated by 1.3 kilometres},
  journal = {Nature},
  volume = {526},
  pages = {682--686},
  year = {2015},
  doi = {10.1038/nature15759},
}

@article{Pompili2021,
  author = {Pompili, M. and Hermans, S. L. N. and Baier, S. and Beukers, H. K. C. and Humphreys, P. C. and Schouten, R. N. and Vermeulen, R. F. L. and Tiggelman, M. J. and dos Santos Martins, L. and Dirkse, B. and Wehner, S. and Hanson, R.},
  title = {Realization of a multinode quantum network of remote solid-state qubits},
  journal = {Science},
  volume = {372},
  pages = {259--264},
  year = {2021},
  doi = {10.1126/science.abg1919},
}

@article{Rondin2014,
  author = {Rondin, L. and Tetienne, J.-P. and Hingant, T. and Roch, J.-F. and Maletinsky, P. and Jacques, V.},
  title = {Magnetometry with nitrogen-vacancy defects in diamond},
  journal = {Rep. Prog. Phys.},
  volume = {77},
  pages = {056503},
  year = {2014},
  doi = {10.1088/0034-4885/77/5/056503},
}

@article{Barry2020,
  author = {Barry, John F. and Schloss, Jennifer M. and Bauch, Erik and Turner, Matthew J. and Hart, Connor A. and Pham, Linh M. and Walsworth, Ronald L.},
  title = {Sensitivity optimization for {NV}-diamond magnetometry},
  journal = {Rev. Mod. Phys.},
  volume = {92},
  pages = {015004},
  year = {2020},
  doi = {10.1103/RevModPhys.92.015004},
}

@article{Hauf2011,
  author = {Hauf, M. V. and Grotz, B. and Naydenov, B. and Dankerl, M. and Pezzagna, S. and Meijer, J. and Jelezko, F. and Wrachtrup, J. and Stutzmann, M. and Reinhard, F. and Garrido, J. A.},
  title = {Chemical control of the charge state of nitrogen-vacancy centers in diamond},
  journal = {Phys. Rev. B},
  volume = {83},
  pages = {081304},
  year = {2011},
  doi = {10.1103/PhysRevB.83.081304},
}

@article{Fu2010,
  author = {Fu, K.-M. C. and Santori, C. and Barclay, P. E. and Beausoleil, R. G.},
  title = {Conversion of neutral nitrogen-vacancy centers to negatively charged nitrogen-vacancy centers through selective oxidation},
  journal = {Appl. Phys. Lett.},
  volume = {96},
  pages = {121907},
  year = {2010},
  doi = {10.1063/1.3364135},
}

@article{Grotz2012,
  author = {Grotz, Bernhard and Hauf, Moritz V. and Dankerl, Markus and Naydenov, Boris and Pezzagna, S{\'e}bastien and Meijer, Jan and Jelezko, Fedor and Wrachtrup, J{\"o}rg and Stutzmann, Martin and Reinhard, Friedemann and Garrido, Jose A.},
  title = {Charge state manipulation of qubits in diamond},
  journal = {Nat. Commun.},
  volume = {3},
  pages = {729},
  year = {2012},
  doi = {10.1038/ncomms1729},
}

@article{Doi2014,
  author = {Doi, Y. and Makino, T. and Kato, H. and Takeuchi, D. and Ogura, M. and Okushi, H. and Morishita, H. and Tashima, T. and Miwa, S. and Yamasaki, S. and Neumann, P. and Wrachtrup, J. and Suzuki, Y. and Mizuochi, N.},
  title = {Deterministic electrical charge-state initialization of single nitrogen-vacancy center in diamond},
  journal = {Phys. Rev. X},
  volume = {4},
  pages = {011057},
  year = {2014},
  doi = {10.1103/PhysRevX.4.011057},
}

@article{Schreyvogel2015,
  author = {Schreyvogel, C. and Polyakov, V. and Wunderlich, R. and Meijer, J. and Nebel, C. E.},
  title = {Active charge state control of single {NV} centres in diamond by in-plane {Al}-{Schottky} junctions},
  journal = {Sci. Rep.},
  volume = {5},
  pages = {12160},
  year = {2015},
  doi = {10.1038/srep12160},
}

@article{Shimizu2023,
  author = {Shimizu, M. and Makino, T. and Kato, H. and Fujiwara, M. and Ogura, M. and Mizuochi, N. and Hatano, M.},
  title = {Charge states of nitrogen-vacancy centers in {Fermi} level controlled diamond n-i-n junctions},
  journal = {J. Appl. Phys.},
  volume = {133},
  pages = {214401},
  year = {2023},
  doi = {10.1063/5.0148921},
}

@article{GrootBerning2014,
  author = {Groot-Berning, Karin and Raatz, Nicole and Dobrinets, Inga and Lesik, Margarita and Spinicelli, Piernicola and Tallaire, Alexandre and Achard, Jocelyn and Jacques, Vincent and Roch, Jean-Fran{\c c}ois and Zaitsev, Alexander M. and Meijer, Jan and Pezzagna, S{\'e}bastien},
  title = {Passive charge state control of nitrogen-vacancy centres in diamond using phosphorous and boron doping},
  journal = {Phys. Status Solidi A},
  volume = {211},
  pages = {2268--2273},
  year = {2014},
  doi = {10.1002/pssa.201431308},
}

@article{Rose2018,
  author = {Rose, B. C. and Huang, D. and Zhang, Z.-H. and Stevenson, P. and Tyryshkin, A. M. and Sangtawesin, S. and Srinivasan, S. and Loudin, L. and Markham, M. L. and Edmonds, A. M. and Twitchen, D. J. and Lyon, S. A. and de Leon, N. P.},
  title = {Observation of an environmentally insensitive solid-state spin defect in diamond},
  journal = {Science},
  volume = {361},
  pages = {60--63},
  year = {2018},
  doi = {10.1126/science.aao0290},
}

@article{Kageura2022,
  author = {Kageura, Taisuke and Sasama, Yosuke and Shinei, Chikara and Teraji, Tokuyuki and Yamada, Keisuke and Onoda, Shinobu and Yamaguchi, Takahide},
  title = {Charge stability of shallow single nitrogen-vacancy centers in lightly boron-doped diamond},
  journal = {Carbon},
  volume = {192},
  pages = {473--481},
  year = {2022},
  doi = {10.1016/j.carbon.2022.01.027},
}

@article{Zhao2026BoronNV,
  author = {Zhao, Gengyou and Tang, Kun and Feng, Bo
            and Gu, Liangxue and Wang, Chuang
            and Wang, Cheng and Wu, Xiaolong
            and Zhu, Shunming and Ye, Jiandong
            and Gu, Shulin},
  title = {The regulation of the crystalline,
           fluorescence, and spin properties of
           nitrogen-doped diamond films by
           controlling boron precursors},
  journal = {Diamond and Related Materials},
  volume = {169},
  pages = {114212},
  year = {2026},
  doi = {10.1016/j.diamond.2026.114212},
}

@article{Broadway2018,
  author = {Broadway, D. A. and Dontschuk, N. and Tsai, A. and Lillie, S. E. and Lew, C. T.-K. and McCallum, J. C. and Johnson, B. C. and Doherty, M. W. and Stacey, A. and Hollenberg, L. C. L. and Tetienne, J.-P.},
  title = {Spatial mapping of band bending in semiconductor devices using in situ quantum sensors},
  journal = {Nat. Electron.},
  volume = {1},
  pages = {502--507},
  year = {2018},
  doi = {10.1038/s41928-018-0130-0},
}

@article{Kageura2024,
  author = {Kageura, Taisuke and Sasama, Yosuke and Yamada, Keisuke and Kimura, Kosuke and Onoda, Shinobu and Yamaguchi, Takahide},
  title = {Surface transfer doping of hydrogen-terminated diamond probed by shallow nitrogen-vacancy centers},
  journal = {Carbon},
  volume = {229},
  pages = {119404},
  year = {2024},
  doi = {10.1016/j.carbon.2024.119404},
}

@article{Macpherson2015,
  author = {Macpherson, Julie V.},
  title = {A practical guide to using boron doped diamond in electrochemical research},
  journal = {Phys. Chem. Chem. Phys.},
  volume = {17},
  pages = {2935--2949},
  year = {2015},
  doi = {10.1039/C4CP04022H},
}

@article{Yang2019,
  author = {Yang, Nianjun and Yu, Siyu and Macpherson, Julie V. and Einaga, Yasuaki and Zhao, Hongying and Zhao, Guohua and Swain, Greg M. and Jiang, Xin},
  title = {Conductive diamond: synthesis, properties, and electrochemical applications},
  journal = {Chem. Soc. Rev.},
  volume = {48},
  pages = {157--204},
  year = {2019},
  doi = {10.1039/C7CS00757D},
}

@article{Klein2007,
  author = {Klein, T. and Achatz, P. and Kacmarcik, J. and Marcenat, C. and Gustafsson, F. and Marcus, J. and Bustarret, E. and Pernot, J. and Omnes, F. and Sernelius, Bo E. and Persson, C. and Ferreira da Silva, A. and Cytermann, C.},
  title = {{M}etal--insulator transition and superconductivity in boron-doped diamond},
  journal = {Phys. Rev. B},
  volume = {75},
  pages = {165313},
  year = {2007},
  doi = {10.1103/PhysRevB.75.165313},
}

@article{Borst1996,
  author = {Borst, T. H. and Weis, O.},
  title = {Boron-doped homoepitaxial diamond layers: Fabrication, characterization, and electronic applications},
  journal = {Phys. Status Solidi A},
  volume = {154},
  pages = {423--444},
  year = {1996},
  doi = {10.1002/pssa.2211540130},
}

@article{Lagrange1998,
  author = {Lagrange, J.-P. and Deneuville, A. and Gheeraert, E.},
  title = {Activation energy in low compensated homoepitaxial boron-doped diamond films},
  journal = {Diam. Relat. Mater.},
  volume = {7},
  pages = {1390--1393},
  year = {1998},
  doi = {10.1016/S0925-9635(98)00225-8},
}

@article{Ekimov2004,
  author = {Ekimov, E. A. and Sidorov, V. A. and Bauer, E. D. and Mel'nik, N. N. and Curro, N. J. and Thompson, J. D. and Stishov, S. M.},
  title = {Superconductivity in diamond},
  journal = {Nature},
  volume = {428},
  pages = {542--545},
  year = {2004},
  doi = {10.1038/nature02449},
}

@article{Takano2004,
  author = {Takano, Y. and Nagao, M. and Sakaguchi, I. and Tachiki, M. and Hatano, T. and Kobayashi, K. and Umezawa, H. and Kawarada, H.},
  title = {Superconductivity in diamond thin films well above liquid helium temperature},
  journal = {Appl. Phys. Lett.},
  volume = {85},
  pages = {2851--2853},
  year = {2004},
  doi = {10.1063/1.1802389},
}

@article{Oing2021,
  author = {Oing, D. and Ney, M. and Bendt, G. and Schulz, S. and Geller, M. and W{\"o}hrl, N. and Lorke, A.},
  title = {A monolithic, back-gated diamond field-effect transistor for tunable color centers},
  journal = {Diam. Relat. Mater.},
  volume = {119},
  pages = {108597},
  year = {2021},
  doi = {10.1016/j.diamond.2021.108597},
}

@article{Krecmarova2021,
  author = {Krecmarova, Marie and Gulka, Michal and Vandenryt, Thijs and Hruby, Jaroslav and Fekete, Ladislav and Hubik, Pavel and Taylor, Andrew and Mortet, Vincent and Thoelen, Ronald and Bourgeois, Emilie and Nesladek, Milos},
  title = {A label-free diamond microfluidic {DNA} sensor based on active nitrogen-vacancy center charge state control},
  journal = {ACS Appl. Mater. Interfaces},
  volume = {13},
  pages = {18500--18510},
  year = {2021},
  doi = {10.1021/acsami.1c01118},
}

@article{Ohkuma2026Coherent,
  author = {Ohkuma, Masahiro and Kimura, Eikichi and Lee, Eunsang
             and Matsumoto, Ryo and Ohyama, Shumpei and Tsuchiya, Saki
             and Lim, Harim and Lee, Yong Soo and Takano, Yoshihiko
             and Lee, Junghyun and Arai, Keigo},
  title = {Coherent control of solid-state defect spins via patterned
             boron-doped diamond circuit},
  journal = {Appl. Phys. Lett.},
  volume = {129},
  number = {6},
  pages = {064004},
  year = {2026},
  doi = {10.1063/5.0336267},
}

@misc{Ohkuma2026Integrated,
  author = {Masahiro Ohkuma and Ryo Matsumoto and Shintaro Adachi
                   and Shinobu Onoda and Kensei Terashima and Takao Watanabe
                   and Yoshihiko Takano and Keigo Arai},
  title = {Diamond quantum-sensing platform with integrated
                   boron-doped diamond microwave antenna and thermometer},
  year = {2026},
  eprint = {2608.27160},
  archiveprefix = {arXiv},
  primaryclass = {physics.app-ph},
  doi = {10.48550/arXiv.2608.27160},
}

@article{Farrer1969,
  author = {Farrer, R. G.},
  title = {On the substitutional nitrogen donor in diamond},
  journal = {Solid State Commun.},
  volume = {7},
  pages = {685--688},
  year = {1969},
  doi = {10.1016/0038-1098(69)90593-6},
}

@article{Collins2002,
  author = {Collins, A. T.},
  title = {The {Fermi} level in diamond},
  journal = {J. Phys.: Condens. Matter},
  volume = {14},
  pages = {3743--3750},
  year = {2002},
  doi = {10.1088/0953-8984/14/14/307},
}

@book{Sze2007,
  author = {Sze, S. M. and Ng, Kwok K.},
  title = {Physics of Semiconductor Devices},
  edition = {3rd},
  publisher = {Wiley},
  address = {Hoboken, NJ},
  year = {2007},
  doi = {10.1002/0470068329},
}

@book{ShklovskiiEfros1984,
  author = {Shklovskii, B. I. and Efros, A. L.},
  title = {Electronic Properties of Doped Semiconductors},
  publisher = {Springer},
  address = {Berlin},
  year = {1984},
  doi = {10.1007/978-3-662-02403-4},
}

@book{Mott1990,
  author = {Mott, N. F.},
  title = {Metal--Insulator Transitions},
  edition = {2nd},
  publisher = {Taylor \& Francis},
  address = {London},
  year = {1990},
  doi = {10.1201/b12795},
}

@misc{SupplementalMaterial,
  note = {See Supplemental Material for the finite-volume discretization, numerical tolerances, substrate-length
  and mesh-convergence tables, high-density model checks, screening
  definitions, and interface-sensitivity data.},
}

@article{Aslam2013,
  author = {Aslam, N. and Waldherr, G. and Neumann, P. and Jelezko, F. and Wrachtrup, J.},
  title = {Photo-induced ionization dynamics of the nitrogen vacancy defect in diamond investigated by single-shot charge state detection},
  journal = {New J. Phys.},
  volume = {15},
  pages = {013064},
  year = {2013},
  doi = {10.1088/1367-2630/15/1/013064},
}

@article{Deak2014,
  author = {De{\'a}k, Peter and Aradi, B{\'a}lint and Kaviani, Moloud and Frauenheim, Thomas and Gali, Adam},
  title = {Formation of {NV} centers in diamond: A theoretical study based on calculated transitions and migration of nitrogen and vacancy related defects},
  journal = {Phys. Rev. B},
  volume = {89},
  pages = {075203},
  year = {2014},
  doi = {10.1103/PhysRevB.89.075203},
}

@misc{Kimura2026Universal,
  author = {Kimura, Eikichi and Rathi, Ashutosh and Tsuchiya, Saki
                   and Mansouri, El Mustapha and Mandi{\'c}, Sara
                   and Saikia, Kristi Kanya and Oshimi, Keisuke
                   and Fujiwara, Masazumi and Arai, Keigo},
  title = {Universal scaling of electrochemical information transfer
                   at solid-liquid interfaces},
  year = {2026},
  eprint = {2607.22190},
  archiveprefix = {arXiv},
  primaryclass = {physics.app-ph},
  doi = {10.48550/arXiv.2607.22190},
}

@article{Kawano2010,
  author = {Kawano, Akihiro and Ishiwata, Hitoshi and Iriyama, Shingo
            and Okada, Ryosuke and Yamaguchi, Takahide
            and Takano, Yoshihiko and Kawarada, Hiroshi},
  title = {Superconductor-to-insulator transition in boron-doped diamond films
           grown using chemical vapor deposition},
  journal = {Phys. Rev. B},
  volume = {82},
  number = {8},
  pages = {085318},
  year = {2010},
  doi = {10.1103/PhysRevB.82.085318},
}

@dataset{Koga2026Data,
  author = {Koga, Masato and Cho, Haizen and Mansouri, EL Mustapha and Kimura, Eikichi and Matsumoto, Ryo and Takano, Yoshihiko and Arai, Keigo},
  title = {Data and analysis code for {Electrostatic screening and nitrogen-vacancy charge states in boron-doped/nitrogen-doped diamond junctions}},
  year = {2026},
  publisher = {Zenodo},
  doi = {10.5281/zenodo.23018029},
}

\clearpage
\setlength{\dblfloatsep}{8pt plus 2pt minus 2pt}
\setcounter{section}{0}
\setcounter{subsection}{0}
\setcounter{equation}{0}
\setcounter{table}{0}
\setcounter{figure}{0}
\renewcommand{\theequation}{S\arabic{equation}}
\renewcommand{\thetable}{S\arabic{table}}
\renewcommand{\thefigure}{S\arabic{figure}}
\renewcommand{\theHsection}{supp.\arabic{section}}
\renewcommand{\theHequation}{supp.\arabic{equation}}
\renewcommand{\theHtable}{supp.\arabic{table}}
\renewcommand{\theHfigure}{supp.\arabic{figure}}
\setlength{\textheight}{\dimexpr\textheight-25pt\relax}
\addtolength{\footskip}{25pt}
\onecolumngrid
\phantomsection
\pdfbookmark[0]{Supplemental Material}{supplemental}
\begin{center}
{\large\bfseries Supplemental Material for ``Electrostatic screening and nitrogen-vacancy charge states in boron-doped/nitrogen-doped diamond junctions''\par}
\vspace{9pt}
Masato Koga\textsuperscript{1}, Haizen Cho\textsuperscript{1}, El Mustapha Mansouri\textsuperscript{1}, Eikichi Kimura\textsuperscript{1}, Ryo Matsumoto\textsuperscript{2}, Yoshihiko Takano\textsuperscript{2}, and Keigo Arai\textsuperscript{1,*}\par
\vspace{3pt}
{\small\itshape
\textsuperscript{1}School of Engineering, Institute of Science Tokyo, 4259 Nagatsuta-cho, Midori-ku, Yokohama,
Kanagawa 226-8501, Japan\\
\textsuperscript{2}National Institute for Materials Science, Tsukuba, Ibaraki, Japan\par}
\vspace{3pt}
{\footnotesize\textsuperscript{*}\href{mailto:arai.k.835f@m.isct.ac.jp}{arai.k.835f@m.isct.ac.jp}\par}
\vspace{9pt}
\begin{minipage}{0.88\textwidth}
\small

This Supplemental Material specifies the dopant and interfacial profiles,
finite-volume discretization, nonlinear solution criteria, substrate-length
and mesh-convergence tests, high-density BDD checks, screening definitions,
and parameter sensitivity data used in the main article. 

\end{minipage}
\end{center}
\vspace{6pt}
\twocolumngrid

\section{Electrostatic profiles and charge model}

The coordinate origin is the nominal BDD/substrate interface.  The
nitrogen-doped substrate occupies $-L_{\mathrm{sub}}\leq z<0$, whereas the BDD
film occupies $0<z\leq t_{\mathrm{BDD}}$, with
$t_{\mathrm{BDD}}=1\ \mu\mathrm{m}$.  The active dopant concentrations are
represented by:
\begin{equation}
\begin{aligned}
s_D(z)&=\frac{1-\tanh(z/w)}{2},\\
N_D(z)&=N_{D0}s_D(z),\\
N_A(z)&=N_{B0}[1-s_D(z)].
\end{aligned}
\label{eq:dopants_si}
\end{equation}
The hyperbolic-tangent profile interpolates between the
bulk dopant concentrations over a scale set by $w$;
it is not intended to specify a unique microscopic
dopant distribution. The positions where $s_D=0.9$ and 0.1 are
separated by $2\,\mathrm{atanh}(0.8)w=2.1972w$.
Unless explicitly varied, $w=8$ nm.
Here $N_{B0}$ denotes the active boron concentration,
reported as $N_B$ in $\mathrm{cm}^{-3}$ throughout this work.
The active substitutional nitrogen concentration $N_{D0}$
is reported as $[\mathrm{N}_{\mathrm{s}}]$ in ppm, with
$1\,\mathrm{ppm}$ corresponding to
$1.76\times10^{17}\,\mathrm{cm}^{-3}$.

The fixed interface sheet charge is represented by
\begin{equation}
\rho_{\mathrm{int}}(z)=
\frac{Q_{\mathrm{int}}}{\sqrt{2\pi}\sigma_{\mathrm{int}}}
\exp\left(-\frac{z^2}{2\sigma_{\mathrm{int}}^2}\right),
\qquad \sigma_{\mathrm{int}}=2\ \mathrm{nm}.
\label{eq:interface_si}
\end{equation}
On each numerical grid, the discrete profile is renormalized such that its
finite-volume integral is exactly $Q_{\mathrm{int}}$.  The sensitivity range
is $|Q_{\mathrm{int}}|/q\leq10^{12}\,\mathrm{cm^{-2}}$.

The band energies are stored in electron volts.  For an electrostatic potential
in volts, the numerical band shifts are
\begin{equation}
E_C(z)=E_{C0}-\psi(z),\qquad E_V(z)=E_C(z)-E_g.
\label{eq:bands_si}
\end{equation}
This is equivalent to the energy shift $-q\psi$ when the energies are expressed
in joules.  The nitrogen donor and boron acceptor levels are
$E_D=E_C-1.7$ eV and $E_A=E_V+0.37$ eV.

The complete Fermi--Dirac integral used in the high-density calculation is
normalized as
\begin{equation}
F_j(\eta)=\frac{1}{\Gamma(j+1)}
\int_0^\infty\frac{x^j}{1+\exp(x-\eta)}\,dx,
\label{eq:fd_si}
\end{equation}
so $F_j(\eta)\rightarrow\exp(\eta)$ in the nondegenerate limit.  The
high-density model uses $p=N_VF_{1/2}[(E_V-E_F)/k_BT]$,
$n=N_CF_{1/2}[(E_F-E_C)/k_BT]$, and $N_A^-=N_A(z)$.  
In the high-density model, complete acceptor ionization
is assumed throughout the graded boron profile,
including its extension into the nitrogen-doped region.
This avoids introducing an abrupt change in the
ionization treatment at the nominal interface.
Nitrogen donors retain the incomplete-ionization
model of Eq.~(5) of the main text.

The NV occupation formula in the main paper treats the stated
NV$^{0/-}$ charge-transition level as an effective finite-temperature
free-energy level.
For a bare transition energy $E_t$ and explicit state degeneracies $g_0$ and
$g_-$,
\begin{equation}
P(\mathrm{NV}^-)=
\left[1+\frac{g_0}{g_-}\exp\left(\frac{E_t-E_F}{k_BT}\right)\right]^{-1}.
\label{eq:nvdeg_si}
\end{equation}
Therefore, the half-occupation condition is shifted by
$k_BT\ln(g_0/g_-)$ relative to $E_t$. In the main calculation, this term is
absorbed into the effective transition level and its 2.6--2.8-eV range.

\section{Finite-volume discretization and nonlinear solution}

Let $z_i$ denote the grid nodes, and let
$\Delta z_i=(z_{i+1}-z_{i-1})/2$ denote the width of an interior control volume.
For a constant dielectric permittivity, the integrated Poisson residual is
\begin{equation}
R_i=\varepsilon\left[
\frac{\psi_{i+1}-\psi_i}{z_{i+1}-z_i}
-\frac{\psi_i-\psi_{i-1}}{z_i-z_{i-1}}
\right]+\rho_i\Delta z_i.
\label{eq:fv_si}
\end{equation}
The unknown vector contains every nodal potential and the global Fermi level.
The remaining equations are as follows:
\begin{align}
\frac{\psi_1-\psi_0}{z_1-z_0}&=0,
\label{eq:leftbc_si}\\
\psi(z=t_{\mathrm{BDD}})&=0,
\label{eq:gauge_si}\\
\sum_{i=1}^{N-2}\rho_i\Delta z_i&=0.
\label{eq:neutral_si}
\end{align}
Equations \eqref{eq:fv_si}--\eqref{eq:neutral_si} imply a zero electric field
at the BDD surface up to the discrete Gauss-law error; this field was not
imposed as an additional condition.

The default mesh has a 2-nm spacing from $z=-2\ \mu$m through the BDD,
20-nm spacing between $-20$ and $-2\ \mu$m, and 1-$\mu$m spacing farther into
the substrate.  The screening map uses a 0.5-nm near-interface and BDD spacing.
Mesh spacings of 2, 1, 0.5, and 0.25 nm are tested for
$(N_B,[\mathrm N_s])=(10^{16},0.01)$,
$(3.7276\times10^{19},100)$, and
$(10^{21},100)$. For these
three cases, the 0.5-to-0.25-nm differences are below
0.005 in the screening fraction and below the larger
of 0.1 nm and 2\% in screening depth.
The high-density BDD calculations using the Fermi--Dirac
model instead employ a 0.05-nm mesh within the first
20 nm of the BDD region to resolve subnanometer screening. The shallow-depth test described in Sec. IV refines the near-interface
mesh on both sides of the junction.  The grid always contains a node at $z=0$.

The coupled equations were solved using a damped Newton method. Continuation
factors of 0, 0.15, 0.40, 0.70, and 1.0 multiply the mobile-carrier response
during initialization.  A solution is accepted when the scaled nonlinear
residual is below $3\times10^{-8}$, the largest change in the potential is below
$2\times10^{-8}$ V, and the change in $E_F$ is below
$2\times10^{-9}$ eV. Post-solution checks require the global-charge and Gauss-law closure
errors to be below $5\times10^{-8}~\mathrm{C\,m^{-2}}$ and the maximum
scaled local flux residual to be below $2\times10^{-7}$. The local residual is
Eq. \eqref{eq:fv_si} divided by
$q\max(N_A,N_D)\Delta z_i$.  Nondegenerate solutions also require
$E_F-E_V\geq3k_BT$ in the BDD and a Boltzmann/Fermi--Dirac hole-density
difference below 10\%.

\section{Neutral-bulk screening and substrate-length convergence}

At 1 ppm of nitrogen, the near-interface
ionized-donor region extends over
approximately 100 nm, whereas the deep-donor
freeze-out produces a much longer
neutral-bulk screening tail.
This separation makes convergence of
the full electrostatic potential more
demanding than convergence of the
local NV charge-state boundary.

To estimate the asymptotic screening length of the weak
neutral-bulk tail, we linearize the charge response about the
charge-neutral bulk state. In neutral nitrogen-doped diamond,
$n_0=N_D^+$ to leading order. For a small positive change in $\psi$, the electron density
increases and the ionized-donor density decreases by the
same amount to first order, giving
\begin{equation}
\frac{d\rho}{d\psi}\simeq-\frac{2q^2n_0}{k_BT},\qquad
\lambda_\infty=\sqrt{\frac{\varepsilon k_BT}{2q^2n_0}}.
\label{eq:tail_si}
\end{equation}
Here, $k_BT$ is expressed in joules, and the carrier
density and permittivity are evaluated in SI units.
For $N_{D0}=1.76\times10^{17}\,\mathrm{cm^{-3}}$,
$N_C=10^{19}\,\mathrm{cm^{-3}}$,
and $E_C-E_D=1.7$ eV, the bulk solution is
$n_0=4.93\times10^3\,\mathrm{cm^{-3}}$ and
$\lambda_\infty=2.87$ cm.
Hence $n_0/N_{D0}=2.8\times10^{-14}$.

\begin{figure*}[tbp]
\includegraphics[width=\textwidth]{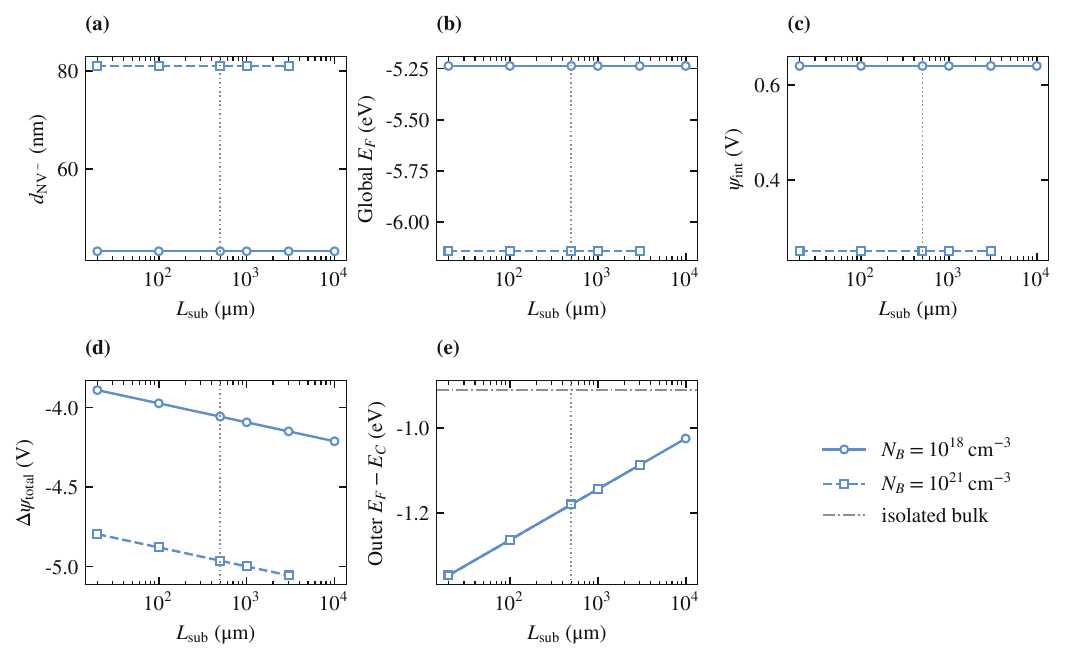}
\caption{Dependence on substrate length at $[\mathrm N_s]=1$ ppm for the
nondegenerate model at $N_B=10^{18}\,\mathrm{cm^{-3}}$ and the adopted
high-density model at $10^{21}\,\mathrm{cm^{-3}}$: (a) NV$^-$ depletion
depth, (b) global Fermi level, (c) interface potential, (d) total
surface-to-surface potential drop, and (e) outer-boundary $E_F-E_C$.  The
vertical line marks 500 $\mu$m; the horizontal line in (e) is the isolated
bulk value.  The high-density 10-mm calculation failed the stated convergence
criteria and is not plotted.}
\label{fig:boundary_si}
\end{figure*}

The NV$^-$ depletion depth shows numerical convergence with
substrate length for both the nondegenerate and high-density
junctions over their respective tested ranges
(Fig.~\ref{fig:boundary_si}).
In the nondegenerate junction, increasing $L_{\mathrm{sub}}$ from 20~$\mu$m to 10~mm changes the NV$^-$ depletion depth
$d_{\mathrm{NV^-}}$ by less than $10^{-3}$~nm. The global Fermi level $E_F$
and interface potential $\psi_{\mathrm{int}}$ also change by less than
$10^{-6}$~eV and $10^{-6}$~V, respectively. For the high-density junction at
$N_B=10^{21}\,\mathrm{cm^{-3}}$, the local depletion depth likewise converges
over the tested range of up to 3~mm. At $L_{\mathrm{sub}}=500~\mu$m, the
nondegenerate and high-density calculations yielded 43.093 and 81.050~nm,
respectively, which are consistent with the longest converged-domain values at the
reported precision.

The total junction potential drop behaved differently. For
$N_B=10^{18}\,\mathrm{cm^{-3}}$, increasing $L_{\mathrm{sub}}$ from 500~$\mu$m
to 10~mm changes the calculated total potential drop from $-4.057$ to
$-4.212$~V. Over the same range, the outer-boundary value of $E_F-E_C$
changes from $-1.180$ to $-1.025$~eV and remains different from the
isolated-bulk value of $-0.911$~eV. Thus, the convergence of the local
NV$^{0/-}$ crossing does not imply the convergence of the total junction
potential drop. We used the independently verified convergence of the local
crossing to establish the reported NV$^-$ depletion depths, without
interpreting the finite-domain total potential drop as an
infinite-substrate quantity.

The unusually long bulk screening length
and the small ratio $n_0/N_{D0}$
also explain why a residual-charge criterion
normalized by the bulk donor density cannot
establish convergence of the remote substrate.
Therefore, the outer-boundary charge and field criteria were used only to select the initial trial domains. Convergence was verified through explicit substrate-length sweeps.
Table~\ref{tab:length_si} summarizes the corresponding convergence data.
The 10-mm high-density calculation did not satisfy the stated convergence
criteria and was therefore excluded; the high-density series was reported
through 3~mm.

\begin{table*}[t]
\caption{Substrate-length convergence at $[\mathrm N_s]=1$ ppm. The BDD
densities are $10^{18}\,\mathrm{cm^{-3}}$ for the nondegenerate model and
$10^{21}\,\mathrm{cm^{-3}}$ for the high-density model. Extra digits are
shown only to demonstrate numerical convergence.}   
\label{tab:length_si}
\begin{ruledtabular}
\begin{tabular}{lrrrrr}
Model & $L_{\mathrm{sub}}$ ($\mu$m) & $d_{\mathrm{NV^-}}$ (nm) &
$E_F$ (eV) & $\psi_{\mathrm{int}}$ (V) & $\Delta\psi_{\mathrm{total}}$ (V) \\
\hline
Nondegenerate & 20 & 43.092697 & -5.237302 & 0.639892 & -3.890776 \\
 & 100 & 43.092696 & -5.237302 & 0.639892 & -3.973941 \\
 & 500 & 43.092696 & -5.237302 & 0.639892 & -4.057403 \\
 & 1000 & 43.092696 & -5.237302 & 0.639892 & -4.093272 \\
 & 3000 & 43.092696 & -5.237302 & 0.639892 & -4.150095 \\
 & 10000 & 43.092696 & -5.237302 & 0.639892 & -4.212350 \\
High density & 20 & 81.049850 & -6.142545 & 0.249898 & -4.795921 \\
 & 100 & 81.049849 & -6.142545 & 0.249898 & -4.879165 \\
 & 500 & 81.049849 & -6.142545 & 0.249898 & -4.962643 \\
 & 1000 & 81.049849 & -6.142545 & 0.249898 & -4.998514 \\
 & 3000 & 81.049849 & -6.142545 & 0.249898 & -5.055338 \\
\end{tabular}
\end{ruledtabular}
\end{table*}

\section{Mesh-convergence tests}

For the nondegenerate isolated-acceptor model, the
approximately 7-nm NV$^-$ depletion depth is stable under
near-interface mesh refinement at $N_B=10^{17}\,\mathrm{cm^{-3}}$
and $[\mathrm N_s]=1$ ppm (Table~\ref{tab:mesh_si}). The substrate length was fixed at 100 $\mu$m, and the
near-interface mesh was refined on both sides.  The 2.0-nm result differs from
the 0.5-nm result by 0.00298 nm, or 0.044\%.  The changes in $E_F$ and $\psi_{\mathrm{int}}$ are below
$3\times10^{-7}$ eV and $3\times10^{-7}$ V, respectively.

\begin{table}[t]
\caption{Mesh convergence of the shallow NV$^-$ depletion depth.}
\label{tab:mesh_si}
\begin{ruledtabular}
\begin{tabular}{rrrrr}
$\Delta z$ (nm) & Nodes & $d_{\mathrm{NV^-}}$ (nm) & $E_F$ (eV) &
$\psi_{\mathrm{int}}$ (V) \\
\hline
2.0 & 2481 & 6.739534 & -5.207472263 & 2.280394371 \\
1.0 & 3981 & 6.736850 & -5.207472426 & 2.280394533 \\
0.5 & 6981 & 6.736558 & -5.207472504 & 2.280394612 \\
\end{tabular}
\end{ruledtabular}
\end{table}

For the high-density BDD calculation at
$N_B=10^{21}\,\mathrm{cm^{-3}}$, refining the first 20 nm of the BDD from
0.10 to 0.05 nm changes the BDD potential drop from
$-0.249898572$ to $-0.249898352$ V and leaves
$d_{\mathrm{NV^-}}=81.049850$ nm at the shown precision.  This test considered only the numerical solution of the continuum model.

\section{VALIDATION AND SENSITIVITY OF THE HIGH-DENSITY BDD MODEL}

The finite-temperature Thomas--Fermi length of the adopted homogeneous BDD
model is
\begin{equation}
\lambda_{\mathrm{TF}}=
\left[\frac{\varepsilon}{q^2(\partial p/\partial\mu)}\right]^{1/2},\qquad
\frac{\partial p}{\partial\mu}=
\frac{N_V}{k_BT}F_{-1/2}(\eta),
\label{eq:tf_si}
\end{equation}
where $\mu\equiv E_V-E_F$ and $\eta=\mu/(k_BT)$.
In Eq.~\eqref{eq:tf_si}, $\mu$ and $k_BT$ are
expressed in joules, with carrier densities
in $\mathrm{m^{-3}}$. For an exponential response, the distance
containing 90\% of the induced charge is
\begin{equation}
d_{90}=\ln(10)\lambda.
\label{eq:d90exp_si}
\end{equation}
At $N_B=10^{21}\,\mathrm{cm^{-3}}$ and $w=0.5$ nm, an exponential fit to the
calculated potential from 1 to 6 nm inside the BDD yields
$\lambda=0.383$ nm. Equation \eqref{eq:tf_si} yielded a value of 0.376 nm.  
The corresponding $d_{90}$ values are 0.882 and 0.866 nm, respectively.

Near the transition between the nondegenerate and high-density modeling regimes, the calculated NV$^-$ depletion
depth is more sensitive to the acceptor-ionization
assumption than to the choice of carrier statistics.
Within the same incomplete-ionization model, replacing
the Boltzmann carrier statistics by Fermi--Dirac
statistics reproduces the Boltzmann results to at least
four significant digits for
$N_B=10^{18}$--$10^{20}\,\mathrm{cm^{-3}}$.
Replacing incomplete with complete acceptor ionization
changes $d_{\mathrm{NV^-}}$ by 8\% at
$N_B=10^{20}\,\mathrm{cm^{-3}}$
(71.78 rather than 66.30 nm).
This quantifies the sensitivity to the ionization model
near the crossover and does not provide experimental
support for complete ionization.

The charge response was varied by scaling $N_C=N_V$ while holding all other
parameters fixed.  Table \ref{tab:bddresponse_si} lists the resulting bulk
chemical potential, differential charge response, and NV$^-$ depletion
depth.  The relative response varies by a factor of 6.3, whereas the depth
varies by 1.27 nm.

\begin{table}[t]
\caption{Sensitivity to the adopted high-density BDD density of states at
$N_B=10^{21}\,\mathrm{cm^{-3}}$ and $[\mathrm N_s]=1$ ppm.}
\label{tab:bddresponse_si}
\begin{ruledtabular}
\begin{tabular}{rrrr}
DOS scale & $E_V-E_F$ (eV) & Relative $\partial p/\partial\mu$ &
$d_{\mathrm{NV^-}}$ (nm) \\
\hline
0.25 & 1.70 & 0.40 & 81.74 \\
0.50 & 1.07 & 0.63 & 81.37 \\
1.00 & 0.67 & 1.00 & 81.05 \\
2.00 & 0.42 & 1.58 & 80.75 \\
4.00 & 0.27 & 2.50 & 80.47 \\
\end{tabular}
\end{ruledtabular}
\end{table}

\section{Screening integrals}

The small-signal charge response was calculated by applying interface sheet
charges $+\delta Q$ and $-\delta Q$, with
$\delta Q/q=0.1\times10^{12}\,\mathrm{cm^{-2}}$, and forming
\begin{equation}
\delta\rho(z)=
\frac{\rho_{\mathrm{mat}}(z;+\delta Q)-\rho_{\mathrm{mat}}(z;-\delta Q)}{2}.
\label{eq:drho_si}
\end{equation}
Here $\rho_{\mathrm{mat}}=q(p-n+N_D^+-N_A^-)$ is the material charge density;
the imposed Gaussian charge was excluded from $\delta\rho$.  The fractions
screened on the nitrogen-doped and BDD sides are
\begin{equation}
f_j=-\frac{\int_j\delta\rho(z)\,dz}{\delta Q},\qquad
f_{\mathrm{dia}}+f_{\mathrm{BDD}}=1,
\label{eq:fractions_si}
\end{equation}
where $z<0$ is the substrate and $z>0$ is BDD. Across all 55 sampled
points, the maximum absolute deviation of
$f_{\mathrm{dia}}+f_{\mathrm{BDD}}$ from unity was
$2.1\times10^{-12}$.

To evaluate these fractions on the numerical grid, we use a
piecewise-constant response in each interior control volume
$C_i=[z_{i-1/2},z_{i+1/2}]$:
\begin{equation}
f_{\mathrm{BDD}}=-\frac{1}{\delta Q}
\sum_i\delta\rho_i\,|C_i\cap(0,t_{\mathrm{BDD}})|.
\end{equation}
The substrate fraction uses $C_i\cap(-L_{\mathrm{sub}},0)$.
The interface cell is split at $z=0$, and the boundary
nodes have zero charge volume,
consistent with the discrete neutrality equation.
The minus sign defines the screening fractions as positive for
induced charge that compensates the applied interface charge.

The BDD 90\% screening depth is the smallest $d>0$ that satisfies
\begin{equation}
\int_0^d|\delta\rho_{\mathrm{BDD}}(z)|\,dz
=0.9\int_0^{t_{\mathrm{BDD}}}
|\delta\rho_{\mathrm{BDD}}(z)|\,dz.
\label{eq:d90_si}
\end{equation}
Taking the absolute value prevents contributions of opposite signs from canceling in Eq. \eqref{eq:d90_si}; $\delta\rho_{\mathrm{BDD}}$ is the material response
restricted to $z>0$. Using the same control volumes, $d$ satisfies:
\begin{equation}
\begin{aligned}
&\sum_i|\delta\rho_i|\,|C_i\cap(0,d)|\\
&\quad=0.9\sum_i|\delta\rho_i|\,|C_i\cap(0,t_{\mathrm{BDD}})|.
\end{aligned}
\end{equation}
The positive-side part of the interface cell is included, and $d$ is
interpolated within the cell where the cumulative charge reaches 90\%.

\section{Interface sensitivity}

The interface grading scale produces the largest
variation in the calculated NV$^-$ depletion depth
in the high-density case
(Table~\ref{tab:interface_si};
see also Fig.~5 of the main manuscript).
At $N_B=10^{21}\,\mathrm{cm^{-3}}$ and
$Q_{\mathrm{int}}=0$, increasing $w$ from
0.5 to 8 nm increases the depletion depth
from 59.15 to 81.05 nm.
In contrast, at $N_B=10^{18}\,\mathrm{cm^{-3}}$
and $w=8$ nm, increasing $Q_{\mathrm{int}}/q$
from $-1.0\times10^{12}$ to
$+1.0\times10^{12}\,\mathrm{cm^{-2}}$
reduces the depth from 49.53 to 32.57 nm.
At $N_B=10^{21}\,\mathrm{cm^{-3}}$,
the depth remains 81.05 nm at the reported
precision over the same charge range.
This negligible high-density response is
specific to the assumed position and width
of the centered Gaussian charge profile.

\begin{table}[t]
\caption{Sensitivity of $d_{\mathrm{NV^-}}$ to interface grading and fixed
interface charge at $[\mathrm N_s]=1$ ppm.}
\label{tab:interface_si}
\scriptsize
\setlength{\tabcolsep}{2.5pt}
\begin{ruledtabular}
\begin{tabular}{llrrr}
Varied quantity & $N_B$ (cm$^{-3}$) & $w$ (nm) &
$Q_{\mathrm{int}}/q$ (cm$^{-2}$) & $d_{\mathrm{NV^-}}$ (nm) \\
\hline
Grading & $10^{21}$ & 0.5 & 0 & 59.15 \\
 & $10^{21}$ & 2.0 & 0 & 61.66 \\
 & $10^{21}$ & 4.0 & 0 & 67.47 \\
 & $10^{21}$ & 8.0 & 0 & 81.05 \\
Fixed charge & $10^{18}$ & 8.0 & $-1.0\times10^{12}$ & 49.53 \\
 & $10^{18}$ & 8.0 & $-0.5\times10^{12}$ & 46.83 \\
 & $10^{18}$ & 8.0 & 0 & 43.09 \\
 & $10^{18}$ & 8.0 & $+0.5\times10^{12}$ & 38.34 \\
 & $10^{18}$ & 8.0 & $+1.0\times10^{12}$ & 32.57 \\
Fixed charge & $10^{21}$ & 8.0 & $-1.0\times10^{12}$ & 81.05 \\
 & $10^{21}$ & 8.0 & 0 & 81.05 \\
 & $10^{21}$ & 8.0 & $+1.0\times10^{12}$ & 81.05 \\
\end{tabular}
\end{ruledtabular}
\end{table}

\end{document}